\documentstyle[12pt]{article}
\title{Bethe Equations "on the Wrong Side of Equator"}
\author{G.P. Pronko \\
{\it Institute for High Energy Physics, Protvino,}\\
{\it Moscow reg., 142284, Russia.}\\
\and 
{\it International Solvay Institute, Brussels, Belgium}\\
\and \\
Yu.G. Stroganov \\
{\it Institute for High Energy Physics, Protvino,}\\
{\it Moscow reg., 142284, Russia.}}
\date{}
\begin{document}
\maketitle

\begin{abstract}
We analyse the famous Baxter's $T-Q$ equations for $XXX$ ($XXZ$) spin
chain and show that apart from its usual polynomial (trigonometric)
solution, which provides the solution of Bethe-Ansatz equations,
there exists also the second solution which should corresponds to
Bethe-Ansatz beyond $N/2$. This second solution of Baxter's equation
plays essential role and together with the first one gives rise to
all fusion relations.   
\end{abstract}

\begin{center}{\bf1. Associated solutions of Bethe-Ansatz equations
\\for $XXX$ - spin chains}
\end{center}

The equations of Bethe-Ansatz in the case of $XXX$ - spin $1/2$ chain
\cite{Be31} could be written in the following form:(see e.g.
\cite{Fa96}) 
\begin{equation}
\label{Bethe}
\biggl(\frac{\lambda_j + i/2}{\lambda_j - i/2}\biggr)^N = \prod_{k\ne
j}^n \frac{\lambda_j - \lambda_k + i}{\lambda_j - \lambda_k - i} 
= -\prod_{k = 1}^n \frac{\lambda_j - \lambda_k + i}{\lambda_j -
\lambda_k - i},
\quad
(j = 1,2, \ldots n),
\end{equation}

where $N$ - the length of the chain (total number of spins) and $n$ -
- the number of parameters $\lambda_j$, which describe the state
vector. 

The total spin of the eigenstate, described by $\lambda_j$ is equal
to $\frac{N}{2}-n$, therefore only states with $n \le N/2$ are
meaningful. One can prove e.g. in the frameworks of QISM (see e.g.
\cite{FaTa81}), that if $n > N/2$, the corresponding
Bethe vector vanishes.

Nevertheless, the solutions of (\ref{Bethe}) with n beyond the
equator $N/2$ do exist and moreover their consideration appears to be
very useful.

In this section we shall prove the following.

\vspace{0.5cm}

\bf {Theorem on extended Bethe-Ansatz for $XXX$ spin chain.}
\rm

For each solution of (\ref{Bethe}) with  $n \le N/2$
there exists the associated one-parametric solution with  
$n^{\ast} = N - n +1 > N/2$.

Proof:

\begin{itemize}
\item
Let us consider the set $\{\lambda_j\}$ which is the solution of
(\ref{Bethe}) with $n \le N/2$.
This set defines the polynomial $Q(\lambda)$
\footnote{In more general situation of inhomogeneous XXZ spin chain
this polynomial was introduced by Baxter \cite{Ba71}}, whose roots
are  $\{\lambda_j\}$:
\begin{equation}
Q(\lambda) = \prod_{j = 1}^{n} (\lambda - \lambda_j).
\end{equation}

The equations (\ref{Bethe}) could be represented in the following
form:
\begin{equation}
(\lambda_j - i/2)^N Q(\lambda_j + i)
 + (\lambda_j + i/2)^N Q(\lambda_j - i) = 0.
 \quad
(j = 1,2, \ldots n).
\end{equation}

wherefrom it follows that the polynomial of the degree $N+n$
\begin{equation}
 (\lambda-i/2)^N Q(\lambda+i) +  (\lambda+i/2)^N Q(\lambda-i)
\end{equation}

vanishes at the roots of polynomial $Q(\lambda)$. For the case of the
simple roots this statement implies the validity of the Baxter
equation for $XXX$ spin chain \cite{Ba72,Ba73}:
\begin{equation}
\label{baxter}
\fbox{$(\lambda-i/2)^N Q(\lambda+i) +  (\lambda+i/2)^N Q(\lambda-i) =
 T(\lambda) Q(\lambda),$}
\end{equation}

where the polynomial $T(\lambda)$ of the degree $N$,
is an eigenvalue of transfer matrix (the trace of monodromy matrix)
for $XXX$ - model.

\item 

Let us divide both sides of (\ref{baxter}) on the product
$Q(\lambda-i)\>Q(\lambda)\>Q(\lambda+i)$
\begin{equation}
\label{vfrac}
\frac{T(\lambda)}{Q(\lambda+i)\>Q(\lambda-i)} =
 R(\lambda-i/2) + R(\lambda+i/2),
\end{equation}

where
\begin{equation}
\label{rr}
R(\lambda) = \frac{\lambda^N}{Q(\lambda-i/2)\>Q(\lambda+i/2)}.
\end{equation}

The rational function $R(\lambda)$ can be presented in the following
form:
\begin{equation}
\label{vvr}
R(\lambda) = \pi(\lambda) + \frac{q_{-}(\lambda)}{Q(\lambda-i/2)}
+ \frac{q_{+}(\lambda)}{Q(\lambda+i/2)},
\end{equation}

where $\pi(\lambda)$, $q_{-}(\lambda)$ and $q_{+}(\lambda)$ are
polynomials, whose degrees satisfy:
\begin{eqnarray}
\label{degrees}
&&\mbox{deg}\>\pi(\lambda) = N-2\>n, \nonumber \\
&&\mbox{deg}\>q_{-}(\lambda) < n, \\
&&\mbox{deg}\>q_{+}(\lambda) < n \nonumber.
\end{eqnarray}

These inequalities will be used in the sequel.

Making use of the representation  (\ref{vvr}) for $R(\lambda)$
let us rewrite equation (\ref{vfrac}).
\begin{eqnarray}
\label{grob}
&&\frac{T(\lambda)}{Q(\lambda+i)\>Q(\lambda-i)} = \pi(\lambda -i/2) +
\pi(\lambda +i/2) + \nonumber \\
&&+ \frac{q_{-}(\lambda -i/2)}{Q(\lambda -i)}
+ \frac{q_{+}(\lambda -i/2)}{Q(\lambda)}
+ \frac{q_{-}(\lambda +i/2)}{Q(\lambda)}
+ \frac{q_{+}(\lambda +i/2)}{Q(\lambda +i)}.
\end{eqnarray}

In the r.h.s. of (\ref{grob}) there are two terms with the
denominator  $Q(\lambda )$:
$$\frac{q_{+}(\lambda -i/2)+q_{-}(\lambda +i/2)}{Q(\lambda)}$$

which are absent in the l.h.s.. The degree of the nominator of this
fraction according to (\ref{degrees}) is less then degree of the
denominator, therefore the two terms should cancel each other, hence
\begin{equation}
\label{q+-}
q_{+}(\lambda) = q(\lambda +i/2),\quad q_{-} = - q(\lambda -i/2). 
\end{equation}

\item
With (\ref{q+-}) the representation for $R(\lambda)$ becomes
\begin{equation}
\label{vr}
R(\lambda) = \pi(\lambda) + \frac{q(\lambda +i/2)}{Q(\lambda+i/2)} -
\frac{q(\lambda -i/2)}{Q(\lambda-i/2)},
\end{equation}

The polynomial $\pi(\lambda)$, as any other, also may be presented as
the finite difference
\begin{equation}
\label{pip}
\pi(\lambda) = \rho (\lambda + i/2) - \rho (\lambda - i/2),
\end{equation}

where $\rho (\lambda)$ is a polynomial of the degree $N-2n+1$
The explicit form of $\rho (\lambda)$ one can obtain e.g. via
binomial polynomials ${\lambda \choose m}, m=0,1,2,...$.
\item
Taking into account (\ref{pip}) we arrive at the following equation
for our rational function $R (\lambda)$:
\begin{equation}
\label{v1}
R(\lambda) \equiv \frac{\lambda^N}{Q(\lambda + i/2) Q(\lambda - i/2)}
=
\frac{P(\lambda + i/2)}{Q(\lambda + i/2)}
- \frac{P(\lambda - i/2)}{Q(\lambda - i/2)}, 
\end{equation}

where $P(\lambda)$ is the last and most important polynomial of this
theorem:
\begin{equation}
P(\lambda) = \rho (\lambda) Q(\lambda) + q(\lambda) 
\end{equation}

Counting the degree of $P(\lambda)$ we obtain $deg\>P(\lambda) =
n^{\ast} = N + 1 - n$.

\item
Now we can get rid of the denominators in (\ref{v1}),
and obtain the fundamental equation:
\begin{equation}
\label{main}
\fbox{$P(\lambda + i/2) Q(\lambda - i/2) - P(\lambda - i/2)
Q(\lambda + i/2) = \lambda^N$.}
\end{equation}

\item
This equation is invariant under substitution $Q \rightarrow -P$,
therefore the roots of the polynomial $P(\lambda)$, which we denote
as $\{\lambda^{\ast}_j\}$ provide the solution of Bethe-Ansatz
equations:
\begin{equation}
\label{Bethep}
\biggl(\frac{\lambda^{\ast}_j + i/2}{\lambda^{\ast}_j - i/2}\biggr)^N
= \prod_{k\ne
j}^{n^{\ast}} \frac{\lambda^{\ast}_j - \lambda^{\ast}_k +
i}{\lambda^{\ast}_j - \lambda^{\ast}_k
- i} \quad (j = 1,2, \ldots n^{\ast}),
\end{equation}

as the roots of $Q(\lambda)$ provide the solution of (\ref{Bethe}).

\item

The polynomial $\rho (\lambda)$ in (\ref{pip}) is defined up to the
arbitrary constant $\alpha$. This implies that the polynomial
$P(\lambda)$, corresponding to $Q(\lambda)$ is actually
one-parametric family:
\begin{equation}
\label{alpha}
P(\lambda ,\alpha) = P(\lambda) + \alpha Q(\lambda),
\end{equation},

with obvious agreement with (\ref{main}).

\bf{QED}.

\end{itemize}

The theorem we just have proven may be illustrated by the concrete
example of the set of polynomials $P$ and $Q$ for the case $N=4$:

\vspace{0.5cm}

\begin{tabular}{|c|c|c|c|c|}
\hline
Number & $S$  & $Q(\lambda)$ & $(2S + 1) i P(\lambda)$
&$T(\lambda)$\\
\hline
1 & 0 & $\lambda^2 + \frac{1}{4}$ & $\lambda^3 + \frac{5}{4} \lambda
+ \alpha (\lambda^2 +
\frac{1}{4})$ & $ 2 \lambda^4 + 3 \lambda^2 - \frac{3}{8}$\\
\hline
2 & 0 & $\lambda^2 - \frac{1}{12}$ & $\lambda^3 + \frac{1}{4} \lambda
+ \alpha (\lambda^2 -
\frac{1}{12})$ & $2 \lambda^4 + 3 \lambda^2 + \frac{13}{8}$\\
\hline
3 & 1 & $\lambda - \frac{1}{2}$ & $\lambda^4 + \lambda^3 + \lambda^2
+ \frac{5}{8} \lambda + 
\alpha (\lambda - \frac{1}{2})$& $2 \lambda^4 + \lambda^2 + 2 \lambda
+ \frac{1}{8}$\\
\hline
4 & 1 & $\lambda + \frac{1}{2}$ & $\lambda^4 - \lambda^3 + \lambda^2
-\frac{5}{8} \lambda + 
\alpha (\lambda + \frac{1}{2})$& $2 \lambda^4 + \lambda^2 - 2 \lambda
+ \frac{1}{8}$\\
\hline
5 & 1 & $\lambda$ & $\lambda^4 - \frac{1}{2} \lambda^2 - \frac{3}{16}
+ \alpha \lambda$& $2 \lambda^4 + \lambda^2 - \frac{7}{8}$\\
\hline
6 & 2 & $1$ & $\lambda^5 + \frac{5}{6} \lambda^3 + \frac{7}{48}
\lambda + \alpha$& $2 \lambda^4 - 3 \lambda^2 + \frac{1}{8}$
\\
\hline
\end{tabular}

\vspace{0.5cm}

Few comments are in order.

\begin{itemize}

\item
The polynomial $Q(\lambda)$ is normalized in such a way that the
coefficient at the highest degree is equal to 1.
Comparing the highest degrees in (\ref{main}), we obtain that the
coefficient at the highest degree of $P(\lambda)$ is equal to
$1/i (N - 2n +1) = 1/ i (2S + 1)$,
where $S$ - is the spin of Bethe state. 

\item

Note, that the existence of one-parametric solution of Bethe
equations "beyond equator" implies that these equations are not
independent. We shall consider the consequences of this fact in a
separate paper.

\item

Let us come back to the equation (\ref{vfrac}). Using the
representation (\ref{v1}) for $R(\lambda)$, we obtain the following
expression for eigenvalues of transfer matrix.
\begin{equation}
\label{t}
T(\lambda) = P(\lambda+i) Q(\lambda-i) - P(\lambda-i) Q(\lambda+i).
\end{equation}

\item
Combining (\ref{main}), and (\ref{t}) we easily obtain the equation: 
\begin{equation}
\label{baxter2}
 (\lambda-i/2)^N P(\lambda+i) +  (\lambda+i/2)^N P(\lambda-i) =
 T(\lambda) P(\lambda),
\end{equation}

similar to Baxter equation (\ref{baxter}).

This means that $P(\lambda)$  may be considered as the second
independent solution of (\ref{baxter}). The arbitrary linear
combination of $Q$ and $P$ is the solution of Baxter equation as
well.

\item
Finally note that the polynomials $Q(\lambda), P(\lambda)$,
satisfying the relation (\ref{main}) are completely different from
the eigenvalues of operators $Q\pm$, which have been constructed in
the series of papers of Bazhanov, Lukyanov and Zamolodchikov
\cite{BaLuZa96-98}. In these papers authors considered the field
theory analogues of some useful construction of lattice integrable
models. The extension of their $Q\pm$ operators for 6-vertex model
\footnote{see also \cite{AnFe96}}
requires external magnetic field which spoils the rotational
invariance of $XXX$ - model.
We intend to give the detailed discussion of the properties of the
solutions for Bethe Ansatz equations with magnetic field in
subsequent publications. Also we intend to discuss the relation of
our associated Bethe system (\ref{Bethe}) and (\ref{Bethep}) with
similar construction of Krichiver, Lipan, Wiegmann and Zabrodin
\cite{KLWZ97}\footnote{authors use a special parameter $\nu$ which
plays the role of the magnetic field}. 

\end{itemize}

\begin{center}{\bf2. Fusion relations for transfer matrices}
\end{center}

 As was emphasised in \cite{BaLuZa96-98,KLWZ97}, the fundamental
 equation
  \begin{equation}
\label{main2}
\fbox{$P(\lambda + i/2) Q(\lambda - i/2) - P(\lambda - i/2)
Q(\lambda + i/2) = \lambda^N$.}
\end{equation}

implies the existence of the class of functional relations known as
fusion relations for transfer matrices (see e.g.\cite{KiRe86}). 

Now we have shown that the fundamental relations (\ref{main2})
follows from Bethe-Ansatz equations, therefore these fusion relations
also arise due to Bethe-Ansatz! 

Let us consider the details of the connection of (\ref{main2})
and fusion relations.

First of all let us define the functions $T_s(\lambda)$ as follows:
\begin{equation}
\label{defts}
\fbox{$T_s(\lambda) = P(\lambda+i(s+\frac{1}{2}))
Q(\lambda-i(s+\frac{1}{2})) -
P(\lambda-i(s+\frac{1}{2})) Q(\lambda+i(s+\frac{1}{2})).$}
\end{equation}

The parameter $s$ may be considered as spin in the auxiliary space
and therefore may take integer or half integer, but generally
speaking the r.h.s. in (\ref{defts}) is well defined for arbitrary
complex $s$.

From this definition immediately follows the equation:
\begin{equation}
\label{anti}
T_{-s-1}(\lambda) = - T_s(\lambda),
\end{equation}

and for particular values of $s$ we have:
\begin{eqnarray}
\label{spec}
&T_{1/2}(\lambda) = T(\lambda),\quad T_{-1/2}(\lambda) =
0,&\nonumber\\
&T_{-1}(\lambda) = - T_0 (\lambda) = -\lambda ^N &
\end{eqnarray}

For the sake of brevity we shall use the following notation:
\begin{equation}
\label{defab}
\Delta (a,b) \equiv P(a) Q(b) - P(b) Q(a)
\end{equation}

The function $\Delta (a,b)$ changes sign while $a \rightarrow b$,
what implies the identity:
\begin{equation}
\Delta (a,b) Q(c) + \Delta (b,c) Q(a) + \Delta (c,a) Q(b) = 0.
\end{equation}

Making use of the definitions (\ref{defts}) and (\ref{defab})
we can rewrite the last equation as follows:
\begin{eqnarray}
&&T_{s_1}(\lambda+i (s_2 - s_3)/3) Q(\lambda+2i (s_3 -s _2)/3) +
\nonumber \\
&&+T_{s_2}(\lambda+i (s_3 - s_1)/3) Q(\lambda+2i (s_1 -s _3)/3) +
\\
&&+T_{s_3}(\lambda+i (s_1 - s_2)/3) Q(\lambda+2i (s_2 -s _1)/3) =0,
\nonumber \\
&& s_1+s_2+s_3+3/2 =0. \nonumber
\end{eqnarray}

Apparently this equation may be considered as generalization of $T-Q$
equation (\ref{baxter}).

Another simple identity:
\begin{equation}
\Delta(a,b) \Delta(c,d) - \Delta(a,c) \Delta(b,d) +
\Delta(a,d) \Delta(b,c) = 0
\end{equation}

leads to the following quadratic relations:
\begin{eqnarray}
&&T_{s_1}(\lambda-i(s_1 +1/2)) T_{s_3-s_2-1/2}(\lambda-i(s_2+s_3+1))
-\nonumber
\\
&&- T_{s_2}(\lambda-i(s_2 +1/2))
T_{s_3-s_1-1/2}(\lambda-i(s_1+s_3+1)) +\\
&&+ T_{s_3}(\lambda-i(s_3 +1/2))
T_{s_2-s_1-1/2}(\lambda-i(s_1+s_2+1)) =
0\nonumber,
\end{eqnarray}

For $s_2=-1$, $s_3=0$, the last equation due to (\ref{anti}) and
(\ref{spec}) may be written as famous fusion relations:
\begin{eqnarray}
&&T_{s}(\lambda-i(s +1/2)) T(\lambda)
=\nonumber
\\
&&= (\lambda + i/2)^N T_{s-\frac{1}{2}}(\lambda-i(s +1))
 +
 (\lambda - i/2)^N T_{s+\frac{1}{2}}(\lambda-i\>s),
\end{eqnarray}

where $T_s(\lambda)$ is the eigenvalue of transfer matrix of quantum
spin $1/2$ and auxiliary spin $s$.

\begin{center}{\bf3. $XXX_{s_q}$ - model}
\end{center}

Now let us consider inverse situation when quantum spin is $s_q$, 
while auxiliary spin 1/2. This situation corresponds to the
$XXX_{s_q}$ spin chain. The above discussion could be easily
generalized for this case.

Indeed, the Bethe ansatz equations have the following form:
(see e.g. \cite{Fa96}) 
\begin{equation}
\label{Bethequ}
\biggl(\frac{\lambda_j + i\>s_q}{\lambda_j - i\>s_q}\biggr)^N =
\prod_{k\ne
j}^n \frac{\lambda_j - \lambda_k + i}{\lambda_j - \lambda_k - i} 
= -\prod_{k = 1}^n \frac{\lambda_j - \lambda_k + i}{\lambda_j -
\lambda_k - i},
\quad
(j = 1,2, \ldots n),
\end{equation}

where the notations are the same as in (\ref{Bethe}). 

Now the set of meaningful solutions $\{\lambda_j\}$ are those for $n
\le s_q\>N$. The eigenvalues of the transfer matrix is given by:
\begin{equation}
\label{tm}
T_{\frac{1}{2}, s_q}(\lambda) = (\lambda + i\>s_q)^N
\prod_{j=1}^{n}\frac{\lambda - \lambda_j - i}{\lambda - \lambda_j}
+(\lambda - i\>s_q)^N
\prod_{j=1}^{n}\frac{\lambda - \lambda_j + i}{\lambda - \lambda_j},
\end{equation}

while $T-Q$ Baxter equations look like:
\begin{equation}
\label{baxtersq}
\fbox{$(\lambda-i s_q)^N Q_{s_q}(\lambda+i) +  (\lambda+i s_q)^N
Q_{s_q}(\lambda-i) =
T_{\frac{1}{2} s_q}(\lambda; s_q) Q_{s_q}(\lambda),$}
\end{equation}

To simplify further consideration we shall limit ourself with the
case $s_q=3/2$.

As we did in the case $s_q=1/2$ we divide both sides of
(\ref{baxtersq}) on the product
$Q(\lambda-i)\>Q(\lambda)\>Q(\lambda+i)$. But now trying to represent
r.h.s. as a finite difference we meet an obstacle do to different
shift of spectral parameter in numerators and denominators of the
fractions. To overcome this difficulty we have to multiply both sides
to the additional multipliers $(\lambda + i/2)^N (\lambda - i/2)^N$
(In general case the number of this auxiliary multipliers is $2 s_q
-1$):
\begin{equation}
\label{vfracsq}
\frac{T(\lambda)(\lambda + i/2)^N (\lambda -
i/2)^N}{Q(\lambda+i)\>Q(\lambda-i)} =
 R(\lambda-i/2) + R(\lambda+i/2),
\end{equation}

where
\begin{equation}
R(\lambda) = \frac{(\lambda - i)^N \lambda^N (\lambda + i
)^N}{Q(\lambda-i/2)\>Q(\lambda+i/2)}.
\end{equation}

Further steps are the same as above and finally we arrive at the
following fundamental relation:
\begin{equation}
\label{main3}
\fbox{$P_{\frac{3}{2}}(\lambda+\frac{i}{2})
Q_{\frac{3}{2}}(\lambda-\frac{i}{2}) -
P_{\frac{3}{2}}(\lambda-\frac{i}{2})
Q_{\frac{3}{2}}(\lambda+\frac{i}{2}) =
(\lambda-i)^N \lambda^N (\lambda+i)^N$,}
\end{equation}

and expression for eigenvalues of transfer matrix $T_{\frac{1}{2}
\frac{3}{2}}(\lambda)$:
\begin{equation}
\label{t32}
T_{\frac{1}{2} \frac{3}{2}}(\lambda) (\lambda+\frac{i}{2})^N
(\lambda-\frac{i}{2})^N =
P_{\frac{3}{2}}(\lambda+i) Q_{\frac{3}{2}}(\lambda-i) -
P_{\frac{3}{2}}(\lambda-i) Q_{\frac{3}{2}}(\lambda+i).
\end{equation}

The illustrative example for the case $s_q = 3/2$, $N=2$:

\vspace{0.5cm}

\begin{tabular}{|c|c|c|c|c|}
\hline
Number & $S$  & $Q(\lambda)$ & $(2S + 1) i P(\lambda)$
&$T(\lambda)$\\
\hline
1 & 0 & $\lambda^3 + \frac{5}{4} \lambda$ & $\lambda^4 + \frac{5}{2}
\lambda^2 +\frac{9}{16}
+ \alpha (\lambda^3 +
\frac{5}{4} \lambda)$ & $ 2 \lambda^2 + \frac{15}{2}$\\
\hline
2 & 1 & $\lambda^2 + \frac{9}{20}$ & $\lambda^5 + \frac{5}{2}
\lambda^3 +\frac{9}{16} \lambda
+ \alpha (\lambda^2 +
\frac{9}{20})$ & $2 \lambda^2 + \frac{11}{2}$\\
\hline
3 & 2 & $\lambda $ & $\lambda^6 + \frac{15}{4} \lambda^4 +
\frac{59}{16} \lambda^2 + \frac{45}{64} + 
\alpha \lambda $& $2 \lambda^2 + \frac{3}{2}$\\
\hline
4 & 3 & $ 1$ & $\lambda^7 + \frac{91}{20} \lambda^5 + \frac{91}{16}
\lambda^3 + \frac{369}{320} \lambda	+
\alpha $& $2 \lambda^2 - \frac{9}{2}$\\
\hline
\end{tabular}

In conclusion of this section we formulate the second theorem, which
generalizes the first one.

\vspace{0.5cm}

\bf {Theorem 2:}
\rm
For each solution of equations (\ref{Bethequ}) with  $n \le s N$,
there exists the one-parametric associated solution with 
$n^{\ast} = 2 s N - n +1 > s N$.

Note that with fundamental relation of the type (\ref{main3}) for
arbitrary $s_q$ we can obtain the rational analogues of all fusion
relations considered in \cite{KiRe86}. 

\begin{center}{\bf4. Trigonometric case - $XXZ$ spin chain}
\end{center}

There we shall consider the Bethe-Ansatz "beyond the equator" for
$XXZ$ spin chain. The general ideas of this generalization are the
same, as in the first section.

We shall use the Baxter's parametrization (see e.g. \cite{Ba73})
for spectral $\phi$ and crossing $\eta$ parameters.
In these notations $T-Q$ Baxter equation looks like:
\begin{equation}
\label{tqtrig}
T(\phi) Q(\phi) = \sin^N(\phi+\eta) Q(\phi-2\eta) +
\sin^N(\phi-\eta) Q(\phi+2\eta).
\end{equation}

Usual q-parameter of the $XXZ$ model is defined by $q=e^{2i\eta}$.

Recall that
\begin{equation}
\label{qtrig}
Q(\phi) = \prod_{j=1}^{n} \sin(\phi-\phi_j)
\end{equation}

is now trigonometric polynomial of the degree $n \le N/2$,
where a set $\{\phi_j\}$ substitutes the set of $\{\lambda_j\}$  in
(\ref{Bethe}), all other notations was introduced in the first
section.

Eigenvalues of the transfer matrix $T(\phi)$ are also trigonometric
polynomial of the degree $N$. Instead of the rational function
$R(\lambda)$ we now have meromorphic function:
\begin{equation}
\label{rtrig}
R(\phi) = \frac{\sin^N \phi}{Q(\phi-\eta)\>Q(\phi+\eta)}.
\end{equation}

The analogue of the decomposition on the primitive fractions in
trigonometric case is the decomposition of (\ref{rtrig}) on to the
primitive functions $1/\sin(\phi-\phi_j \pm \eta)$ for odd $N$ and
$\cot(\phi-\phi_j \pm \eta)$ for even $N$.

Making use of such expansion and taking into account Bethe Ansatz we
obtain the trigonometric analogue of representation (\ref{vr}):
\begin{equation}
\label{vrtrig}
R(\phi) = \pi(\phi) + \frac{q(\phi+\eta)}{Q(\phi+\eta)} -
\frac{q(\phi-\eta)}{Q(\phi-\eta)},
\end{equation}

where $\pi(\phi)$ is the trigonometric polynomial of the degree
$N-2n$, while $deg\>q)\phi) < n$.

Now the construction of the $P(\phi)$ which is the analogue  of
$P(\lambda)$ is reduced to the construction of the trigonometric
polynomial $\rho(\phi)$ satisfying 
\begin{equation}
\label{problem}
\rho(\phi+\eta) -\rho(\phi-\eta) \equiv \pi(\phi).
\end{equation}

In the present paper we shall consider the case of $q$ - parameter is
not the root of unity i.e. $\eta$ is not the rational part of $\pi$.

In this case $\sin(k \eta) \ne 0,\>\>k\in Z$ and so we can use the
following simple formulas:
\begin{eqnarray}
\label{school}
&&\sin(k \phi) = \frac{\cos(k (\phi-\eta))-\cos(k (\phi+\eta))}
{\sin(k \eta)}, \nonumber \\
&&\cos(k \phi) = \frac{\sin(k (\phi+\eta))-\sin(k (\phi-\eta))}
{\sin(k \eta)},
\end{eqnarray}

For odd $N$, the degree of $\pi(\phi)$ is also odd and it may be
decomposed in the harmonics $\cos(k \phi), \sin(k \phi)$ á $k \ne 0$.

In this case the equations (\ref{school}) solve the problem
(\ref{problem}) and $\rho(\phi)$ is the trigonometric polynomial of
the degree $N-2n$. The polynomial:
\begin{equation}
\label{ptr}
P(\phi) \equiv \rho(\phi) Q(\phi) + q(\phi),
\end{equation}

is the second solution of (\ref{tqtrig}). Its degree is $N-n$.

Apparently its decomposition:
\begin{equation}
\label{ptrig}
P(\phi) = const \prod_{j=1}^{n^{\ast}} \sin(\phi-\phi^{\ast}_j),
\end{equation}

where $n^{\ast}=N-n$ gives the solution for trigonometric
Bethe-Ansatz equation

\begin{equation}
\label{Bethetrig}
\biggl(\frac{\sin(\phi_j + \eta)}{\sin(\phi_j - \eta)}\biggr)^N =
\prod_{k\ne j}^{n^{\ast}} \frac{\sin(\phi_j - \phi_k + 2 \eta)}
{\sin(\phi_j - \phi_k - 2 \eta)} \quad
(j = 1,2, \ldots n^{\ast})
\end{equation}

"beyond equator".

For even $N$, the polynomial $\pi(\phi)$ has the zero harmonic and
therefore the solution of (\ref{problem}) acquires term with linear
(nonperiodic) dependence of $\phi$.

As the result we have the following

\bf {Theorem on the associated solution of Baxter equation for $XXZ$
spin chain}
\rm

For odd length of spin chain $N$ the equation (\ref{tqtrig})
has the couple of associated solutions which are the trigonometrical
polynomials of the degrees $n \le N/2$ and  $n^{\ast} = N - n > N/2$.

For even length $N$ one solution is the trigonometrical polynomial
of the degrees $n$ while the second has the form (\ref{ptr}), where
$\rho(\phi)$ contains the linear (nonperiodic) dependence of $\phi$.

\vspace{0.5cm}
For the construction of fusion relation in the case of $XXZ$ - model
it is sufficient to use two main ingredients - the analogues of eqs.
(\ref{main2}) and (\ref{defts}). 

The first one can be extracted from representation for $R(\phi)$:
\begin{equation}
P(\phi + \eta) Q(\phi - \eta) - P(\phi - \eta) Q(\phi + \eta) =
\sin^N \phi.
\end{equation}

The second may be written as follows:
\begin{equation}
T_s(\phi) = P(\phi + (2s +1)\eta) Q(\phi - (2s +1)\eta) - P(\phi -
(2s +1)\eta) Q(\phi +(2s +1)\eta).
\end{equation}

Therefore all the results of Sections 2 and 3 holds true for generic
$XXZ$ spin chain.

{\bf Acknowledgements}		
\noindent
We are grateful to L.D. Faddeev, A.V. Razumov, M.V. Saveliev, S.M.
Sergeev and A.Yu. Volkov for useful discussion.

The special thanks to V.V. Bazhanov who attracted our attention to
the fact of existence of two $Q\pm$ operators for $XXZ$ spin chain
with a magnetic field and initiated the present paper. 

The research was supported in part by RFFR grant 98-01-00070 and
INTAS 96-690.

\end{document}